\documentclass[sigconf]{acmart}
\acmConference[SE 2030]{International Workshop on Software Engineering in 2030}{November 2024}{Porto de Galinhas (Brazil)}
\usepackage{amsmath,amsfonts}
\usepackage{algorithmic}
\usepackage{array}
\usepackage{textcomp}
\usepackage{stfloats}
\usepackage{url}
\usepackage{verbatim}
\usepackage{graphicx}
\usepackage{graphicx, subfigure}
\usepackage{caption}
\usepackage{makecell}
\usepackage{soul}
\usepackage{listings}
\usepackage{multirow}
\usepackage{pifont}
\usepackage{xspace}
\usepackage{epigraph} 
\usepackage[flushleft]{threeparttable}
\usepackage{url}
\usepackage{algorithm2e}
\UseRawInputEncoding
\usepackage[colorinlistoftodos]{todonotes}
\usepackage{listings}
\usepackage{comment}
\usepackage{fontawesome}
\usepackage{booktabs}
\usepackage[english]{babel}
\usepackage{xspace}
\usepackage{hyperref}
\usepackage{enumitem}
\hypersetup{
    colorlinks=true,
    linkcolor=blue,
    anchorcolor=blue,
    filecolor=blue,      
    citecolor=blue,
    urlcolor=blue
}

\usepackage{csquotes}
\renewcommand{\mkbegdispquote}[2]{\itshape}

\definecolor{celestialblue}{rgb}{0.29, 0.59, 0.82}
\definecolor{awesome}{rgb}{0.5, 0.2, 0.3}
\definecolor{coolblack}{rgb}{0.0, 0.18, 0.39}
\definecolor{maroon}{cmyk}{0, 0.87, 0.68, 0.32}
\definecolor{halfgray}{gray}{0.55}
\definecolor{ipython_frame}{RGB}{207, 207, 207}
\definecolor{ipython_bg}{RGB}{247, 247, 247}
\definecolor{ipython_red}{RGB}{186, 33, 33}
\definecolor{ipython_green}{RGB}{0, 128, 0}
\definecolor{ipython_cyan}{RGB}{64, 128, 128}
\definecolor{ipython_purple}{RGB}{170, 34, 255}
\usepackage[strict]{changepage}
  \definecolor{ABlue}{HTML}{127bca}
 \definecolor{LHScolor}{HTML}{555555}
\usepackage{framed}

\definecolor{formalshade}{rgb}{1.0,1.0,1.0}
\definecolor{side}{rgb}{0.0,0.2,0.6}
\usepackage[skins,breakable]{tcolorbox}
\definecolor{Large}{HTML}{696969}
\definecolor{Negligible}{HTML}{D3D3D3}
\definecolor{Medium}{HTML}{808080}
\definecolor{Small}{HTML}{A9A9A9}
\definecolor{backcolour}{rgb}{0.95,0.95,0.92}

\definecolor{chestnut}{rgb}{0.8, 0.36, 0.36}

\definecolor{chestnut}{rgb}{0.8, 0.36, 0.36}

\usepackage[skins,breakable]{tcolorbox}

\begin{document}

\title{The Role of Code Proficiency in the Era of Generative AI}



\author{Gregorio Robles}
\affiliation{%
  \institution{Universidad Rey Juan Carlos}
  \streetaddress{}
  \city{Madrid}
  \state{}
  \country{Spain}}
\email{grex@gsyc.urjc.es}

\author{Christoph Treude}
\affiliation{%
  \institution{Singapore Management University}
  \streetaddress{}
  \city{Singapore}
  \state{Singapore}
  \country{Singapore}}
\email{ctreude@smu.edu.sg}

\author{Jesus M. Gonzalez-Barahona}
\affiliation{%
  \institution{Universidad Rey Juan Carlos}
  \streetaddress{}
  \city{Madrid}
  \state{}
  \country{Spain}}
\email{jgb@gsyc.urjc.es}

\author{Raula Gaikovina Kula}
\affiliation{%
  \institution{Nara Institute of Science and Technology}
  \city{Nara}
  \country{Japan}}
\email{raula-k@is.naist.jp}




\markboth{Journal of \LaTeX\ Class Files,~Vol.~14, No.~8, August~2021}%
{Shell \MakeLowercase{\textit{et al.}}: A Sample Article Using IEEEtran.cls for IEEE Journals}


\begin{abstract}
At the current pace of technological advancements, Generative AI models, including both Large Language Models and Large Multi-modal Models, are becoming integral to the developer workspace. However, challenges emerge due to the 'black box' nature of many of these models, where the processes behind their outputs are not transparent. This position paper advocates for a 'white box' approach to these generative models, emphasizing the necessity of transparency and understanding in AI-generated code to match the proficiency levels of human developers and better enable software maintenance and evolution. We outline a research agenda aimed at investigating the alignment between AI-generated code and developer skills, highlighting the importance of responsibility, security, legal compliance, creativity, and social value in software development. The proposed research questions explore the potential of white-box methodologies to ensure that software remains an inspectable, adaptable, and trustworthy asset in the face of rapid AI integration, setting a course for research that could shape the role of code proficiency into 2030 and beyond.


\end{abstract}

\maketitle

\section{Introduction}
{A}{rtificial} Intelligence, particularly large language models (LLMs), is changing how we think about building and maintaining software~\cite{peng2023impact}.
Some voices predict that in a not-too-distant future it will no longer be necessary to know the ins and outs of a programming language to be able to perform software engineering tasks~\cite{kuhail2024will}. Some others are concerned about the lack of control that this change could have for humans in charge of ensuring software systems work as intended~\cite{kirova2023:ethics}.
From a historical perspective, some of these issues are reminiscent of the discussions of several decades ago, when compilers allowed us to abstract from assembly code and use higher-level languages.
At first, many of the (assembler) programmers of that time did not trust the compilers and checked that the translation was done correctly. As work with high-level languages became more popular, this concern decreased, but the problem of to what extent compiler-produced code reflects the intentions of the programmer remained for a long time~\cite{boyle1999you}.
Nowadays, this is overcome, confidence in code produced by compilers is commonplace, and only in very specific cases is it necessary for humans to work at the assembler level. However, the situation now may be different, because we do not understand how generative AIs produce code with the same detail that we understand how compilers produce code. In addition, the gaps between prompts and the code they produce, and the gap between source code in a high-level programming language and the code produced when compiling it, are very different, much higher in the first case.

As famous physicist Niels Bohr presumably said, predicting is complicated, especially the future.
It may be the case that software engineers of the future will not need programming knowledge to perform their tasks.
We call this case the ``black-box case''. But it could also be that they still need to read and understand the source code in a programming language. We call this case the ``white box case''. And, of course, the future could be (and likely will be) complicated, with a range of situations between these cases, depending on the scenario~\cite{sauvola2024:future-development-ai}.

In the ``pure'' black-box case, software engineers will work at a very high level of abstraction, likely using natural language to transmit specifications to an AI agent, which will in turn produce computer programs ready to execute. They could be directly binary code or produced with the intermediation of some source code in traditional programming languages, but not visible to the engineer. Since specifying complex behaviors in natural languages is prone to misunderstandings, 
it is likely that the desired solution will not be achieved in a single iteration, so the engineer should check the output and reformulate or expand the prompt until the desired software product is achieved. Therefore, the complete process could be similar to a golf game, with successive shots getting us closer and closer to the final goal. Several methodologies have been proposed along these lines, such as, for example, the STL (specify-test loop) approach~\cite{gonzalezbarahona2024:llms-xr}.

In the ``pure'' white-box case, software engineers will likely use AI agents to write source code, but they will be more like companions in a pair programming team than anything else~\cite{robe2022:pair-programming-ai}. The engineer will still read source code in a high-level programming language, touch it when needed, and in general use agents to help write and explain code. Engineers will have access to all implementation details and will need to comprehend them. This case is in the end quite similar to the current situation, but with much more powerful tools helping both to write and to understand source code.

\begin{figure}[t]
    \centering
    \includegraphics[width=1\linewidth]{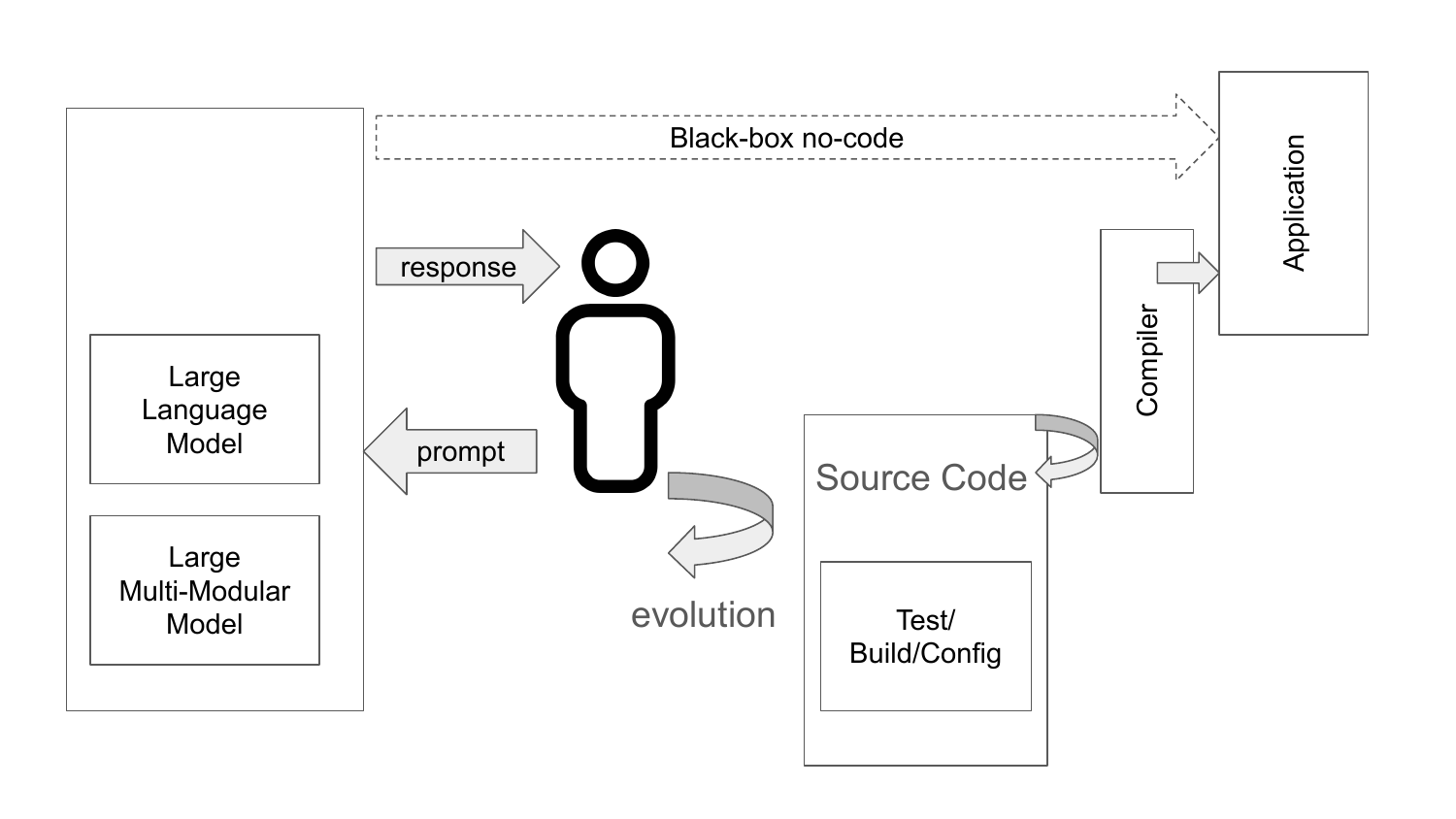}
    \caption{Overview of black-box vs. white-box when using AI agents. The upper part of it shows the black-box case: the software is directly produced by the LLM model, with no human intervention (except for prompting it, and interacting with the resulting software). The lower part of if shows how the white-box case includes humans in the loop, letting them inspect the produced source code.}
    \label{fig:overview}
\end{figure}

We can explore in some more detail the differences between the white-box and black-box cases. Figure \ref{fig:overview} shows a conceptual diagram that highlights the key distinctions between them with respect to how they relate to human software engineers. In a black-box scenario, a human uses a prompt to generate some software. Perhaps there is an intermediary step, where the source code produced by the LLM is compiled into object code. But the central idea is to avoid any human interaction since the prompt is provided up to the moment the software is produced. In contrast, the white-box approach allows a human to inspect the details of the outputs generated by the AI agent. This model positions a human between the LLM and the source code, facilitating multiple iterations of prompting to refine and evolve the desired code output.

Faced with these two extreme scenarios, we consider that usually the white-box approach will be preferred, even when for some scenarios black-box approaches can also be useful and appropriate.

\section{Evolution and trust}

One of our key reasons for favoring a white-box approach lies in the very nature of software development and maintenance.
To remain useful, software needs to evolve~\cite{lehman1996laws}.
It has to be constantly modified because of changes in expectations and requirements by the humans using it and because of changes in the environment (changes in the software system on which the program works, in other components with which it interacts, in the hardware in which it runs, etc.). In a successful software project, at least 80\% of the effort goes to its evolution and maintenance~\cite{chang2001handbook}.
For example, studies show that old software projects, more than 30 years old, with a significant size (in the range of several million lines of code) have a code base with a large proportion of lines written recently (between 30\% and 50\% are less than 5 years old)~\cite{robles2005empirical}.





%

\begin{table*}
\centering
\caption{Four FOSS projects showing to which extent their source code is recent. ``LoC 2024'' offers the lines of code in their most recent version. ``LoC 2024 introduced after 2018'' is the number of LoC in 2024 that have been introduced in 2018 or later (i.e., has been added in the last six years). ``LoC all history'' gives the number of LoC added in all the history of the project. ``Removed LoC all history'' is the LoC that have been removed. LoC stands for Lines of Code (considering comments but not blank lines).}
\begin{tabular}{lrrrr}
\toprule
       & LoC 2024 & LoC 2024 introduced after 2018 & LoC all history & Removed LoC all history \\ \midrule
gimp   & 1,050K & 410K & 2,714K & 1,664K \\
gtk    & 898K & 639K & 1,808K & 910K \\      
http   & 451K & 101K & 893K & 442K \\
wine   & 4.8M & 2.5M & 7.6M & 2.8M \\
        \bottomrule
\end{tabular}
    \label{tab:evolution}
\end{table*}

The software that we run today requires constant updating.
Contrary to what many might intuitively think, software is not usually the result of being created from scratch, and then have just a few specific changes.
On the contrary, software is usually the result of many, very profound changes, many of them performed recently.
In this sense, software differs from texts written in natural language, which are only occasionally modified to keep them current.
In software, although the main aim of a program may remain the same since its beginning, changes are radical, both in what has been added and in the corrections.

In Table~\ref{tab:evolution}, we present concrete evidence illustrating the substantial maintenance effort needed to maintain and evolve the software. The data from a set of Free/Open Source Software (FOSS) projects show that i) although the current size of the codebase exceeds its original dimensions, the volume of code modifications applied (both additions and deletions) surpasses even the final size of the product itself, and ii) that the amount of recent code (i.e., lines included later than 2018) is a significant part of the current body of source code. These observations underscore the fact that code is a dynamic entity, constantly shaped by external and internal bugs, the introduction of new features, and the evolving requirements of users. 

And we know that maintaining software requires knowledge and skills that are different from creating software~\cite{jorgensen2002impact}. They are difficult to exercise with a black-box approach because in many cases even specifying what exactly should be changed is quite a challenge without access to the source code. In those cases, being able to see inside the box, comprehending implementation details, is almost certainly necessary.

Another key reason for the white box case is that engineers will not blindly trust the code produced by AI agents, or that they will find too difficult or time-consuming to interact with until they are sure that the produced code actually performs as they intend. Current generative AI agents are prone to hallucinations~\cite{liu2024:hallucinations-ai-code}, and may produce code with bugs~\cite{tambon2024:bugs-ai-code}. In addition, specifying detailed behaviors in natural language is difficult due to its inherent lack of precision, the possibility of different interpretations, and the challenges of expressing complex actions. We also face the problem that, up to now, models are not explainable. That is, we do not completely understand how they are going to react to specific prompts or how small changes in prompts affect the results produced by a generative model. Furthermore, the issue of reproducibility and non-determinism in AI-generated code adds another layer of complexity, as the same prompt might not always result in the same output, making it challenging to ensure consistent and reliable software behavior.

If we cannot trust the code produced by AI agents at least at the same level as we trust the code produced by humans, we are going to want software engineers to check the code they produce and eventually fix it. This will be fundamental in fields such as safety-critical systems which are subject to specific certification requirements, which usually require that the code has been inspected in detail. But even in less demanding fields, being reasonably sure that the software is going to behave as expected may be challenging without at least the possibility of inspecting the code. 

\section{Advantages of the White-box paradigm}

We argue that adopting a white-box paradigm for software maintenance, evolution, and even initial creation offers several advantages.

a) {\bf Responsibility.}
In some sense, code acts as law~\cite{lessig2000code}.
Software acts, to some extent, as contracts do in the legal world, and nobody would like to sign an opaque contract, although very often only lawyers can understand the legal jargon.
It is not difficult to argue for the importance of software in the daily life of modern society, and while errors in software are inevitable, it is not the same if they are the result of human errors, or a consequence of being no humans verifying that software behaves as intended. A white-box approach, where software can be (and is) inspected and understood by humans, seems like a reasonable requirement for this responsibility.


b) {\bf Security.}
The possibility of working at the source code level allows security audits to be carried out on the created software systems.
The argument in this regard is similar to what open source proponents advocate about the advantages of being able to see the source code: security by inspection versus security by obscurity.
To be able to audit the security of software, the code produced by an LLM should be of high level and understandable by software developers.


c) {\bf Legal compliance and avoidance of bias}
Similarly, if the code cannot be seen, it cannot be certified that the software system does what it should do or, on the contrary, does not do what it should not do.
For the first thing, access to the source code is essential, and that it can be verified, by humans with the help of tools, that it is correct and that it meets certain parameters of quality and stability, among others.
For the second, it is also necessary to be able to access the source code and verify that the software does not contain bias towards people or groups.


d) {\bf Creativity.}
Seeing source code is a trigger for creativity.
Seeing the current state of the source code can be a source of new inspiration, new possibilities, and innovation~\cite{groeneveld2021exploring}.
Creativity is clearly limited in a black-box environment.
In this sense, software development, particularly free software, has been classified as a stigmatic process in which the source code plays the role of signaling.
In other words, just as changes in the environment cause insects such as ants to behave differently, it has been argued that changes in source code have a pulling power for new changes.
That is why, even without embracing free software models, there have been movements by large companies to show the source code of their software, so that people external to these companies can contribute.


e) {\bf Social value.}
In recent years, we have seen a revolution in software development toward social development mechanisms~\cite{de2005seeking}.
Thus, the software gains value through the exchange of opinions, collective supervision, and the possibility of remixing existing source code (forking).
Social software development requires, for optimal development, a white-box environment.
This does not mean that social development cannot be carried out to improve prompt engineering, but this would still be one of the many fields where software development benefits from it.

Our position, therefore, comes to say that software development, and especially its maintenance and evolution, is in constant exchange that creativity is maximized, as well as social value.
It is this exchange that offers the necessary trust - when many other developers have seen it - that allows certain degrees of responsibility to be offered.


\section{Proficiency is key}

Of all the implications of the previous discussion, we will focus on one that we consider of utmost importance: the proficiency in understanding source code. In the white-box scenario, source code generated by LLMs must be not only accessible but also understandable by developers. Of course, developers may be assisted by tools, some of them being AI agents themselves, which help them to check for functional or non-functional properties of the produced code. But they should also be able to understand and reason about the code produced.

In other words, in addition to giving precise instructions about what we want an LLM to generate, questions related to the skills that the developer or development team possesses must also be included as input to the model.
This goes beyond the programming language in which the LLM must generate code, as well as other constraints such as the style guide and the legibility of the source code. In addition, it is fundamental that the code produced by the AI agent is adjusted to the level of knowledge of the software engineers who will interact with it.

For natural languages (English, German, Spanish, etc.), there are frameworks to determine the level of proficiency of a language.
For example, the CEFR, or Common European Framework of Reference for Languages, is an international reference framework that describes levels of proficiency in a foreign language in Europe~\cite{jones2009european}, although there are other frameworks with a similar philosophy in other countries~\cite{nishizawa2022review,liskin1984actfl}.
It is divided into six levels: A1, A2, B1, B2, C1, and C2, from beginner to advanced.
The CEFR is a useful tool to assess student progress, compare levels between different languages, and facilitate student and worker mobility.

If we ask an LLM for a text or description in a foreign language, we would need it to be produced in accordance to our skills in that language if we want to understand it and maybe modify it.
 English Wikipedia, for example, has a Simple English Wikipedia project~\cite{hwang2015aligning}, with requirements such as ``articles should use only the 1,000 most common and basic words in English. They should also use only simple grammar and shorter sentences.''

We consider that it will be necessary as well to specify how we want AI agents to produce their code: which constructs they will use, which level of complexity, and, in general, which elements of the programming language. In this way, we could adapt the output of the LLM to the skills and capabilities of the engineers that should interact with it, in a way that favors its comprehension and makes it easier for them to understand and, if needed, change it. Therefore, it will be important to create proficiency level assessment frameworks for programming languages, just as has been done for natural languages.

This would allow developers to understand, analyze, adapt and modify the code generated by the LLMs, maximizing the advantages of the white-box point of view, as we have argued in the previous paragraph.
So, if a developer has a B2 (intermediate) level of Python, they should be able to ask the LLM for a piece of code that performs an algorithm or corrects a bug at ``Python level B2". LLMs would be trained with these requirements in mind, for example, automatically labeling their training set with the corresponding ``proficiency level". Of course, the LLM can be customized not only in terms of the level of mastery, but also in terms of the style guide, the size and details of comments (and the natural language used for them), for example.

For software projects produced from scratch, the proficiency level of the source code could be defined in advance, given the expected skills of the engineers working on that project. For already existing software projects, an analysis of the level of their source code could be produced, so that further output by AI agents matches that level, which engineers are accustomed to.


\section{Research Agenda}

We outline a research agenda below, presenting specific questions to guide further exploration of white-box vs.~black-box approaches in the era of generative AI and their implications for code proficiency.

\paragraph{Code as a reflection of developer skills}
The inquiry into how the quality of code produced by a program reflects the developer's expertise delves into understanding the effectiveness of automated programming tools in relation to the user's skill level. 
Hence, we ask the first two research questions:
\begin{tcolorbox}[colback=gray!5,colframe=awesome,title=Skills]
\begin{itemize}[leftmargin=*]
    \item How well does the quality of code produced by a program mirror the developer's expertise?
    \item How well does the number of iterations required to produce high-quality code reflect developer skill level?
\end{itemize}
\end{tcolorbox}

\paragraph{Leveraging Competency with AI}
Our next research questions aim to investigate how AI-generated code, when aligned with developer proficiency levels within a white-box framework, impacts various dimensions of software development, including responsibility, security, legal compliance, and the broader socio-technical ecosystem.

\begin{tcolorbox}[colback=gray!5,colframe=awesome,title= Responsibility]
\begin{itemize}[resume,leftmargin=*]
    \item How can a white-box approach to AI-generated code help delineate responsibility for code outcomes when multiple proficiency levels are involved?
    \item In what ways does aligning AI-generated code with developer proficiency levels affect the attribution of responsibility for the code's quality and its outcomes?
\end{itemize}
\end{tcolorbox}

\begin{tcolorbox}[colback=gray!5,colframe=awesome,title= Security]
\begin{itemize}[resume,leftmargin=*]
    \item How does tailoring AI-generated code to match developer proficiency levels within a white-box framework impact the code's resilience against security vulnerabilities?
\end{itemize}
\end{tcolorbox}

\begin{tcolorbox}[colback=gray!5,colframe=awesome,title= Legal]
\begin{itemize}[resume,leftmargin=*]
    \item How can ensuring that AI-generated code is adaptable to various levels of developer proficiency within a white-box approach mitigate legal risks and reduce biases in software?
\end{itemize}
\end{tcolorbox}
\begin{tcolorbox}[colback=gray!5,colframe=awesome,title= Creativity]
\begin{itemize}[resume,leftmargin=*]
    \item Is there a tendency for high-quality code to exhibit greater creativity?
    \item Can AI-generated code lead to a reduction in diverse problem-solving approaches?
\end{itemize}
\end{tcolorbox}

\begin{tcolorbox}[colback=gray!5,colframe=awesome,title= Social]
\begin{itemize}[resume,leftmargin=*]
    \item Do developers have the necessary skills to review automatically generated code effectively?
    \item Does code that reflects the creator's proficiency improve developers' trust and perception, even without human verification?
\end{itemize}
\end{tcolorbox}

\section{Conclusion}

Developers integrating AI into their software development environments is not just a trend but an inevitability, many having already embraced this shift. This paper has outlined various concerns associated with AI in software development and has argued for the critical importance of adopting a white-box approach. By ensuring that source code serves as a transparent measure of proficiency and understanding, we underscore its vital role in the future of software development as we move towards 2030 and beyond.

\bibliographystyle{ACM-Reference-Format}
\bibliography{software}

\end{document}